\documentclass[12pt,preprint]{aastex}
\usepackage{epsfig}
\usepackage{multirow}
\usepackage{graphics}
\usepackage{amsmath, amsthm, amssymb}
\tightenlines

\newcommand \lsim{\mathrel{\rlap{\lower4pt\hbox{\hskip1pt$\sim$}}
    \raise1pt\hbox{$<$}}}
\newcommand \gsim{\mathrel{\rlap{\lower4pt\hbox{\hskip1pt$\sim$}}
    \raise1pt\hbox{$>$}}}

\newcommand     \kms    {\,{\rm km~s}^{-1}}

\newcommand{\beq}{\begin{equation}}
\newcommand{\eeq}{\end{equation}}
\newcommand{\beqa}{\begin{eqnarray}}
\newcommand{\eeqa}{\end{eqnarray}}

\newcommand{\thco}	{^{13}{\rm CO}}
\newcommand{\ceto}	{{\rm C^{18}O}}
\newcommand{\cothree}     {^{13}{\rm CO}}

\newcommand{\gcc}         {\rm g\:cm^{-2}}

\newcommand{\Jtwoone}	{J=2\rightarrow 1}
\newcommand{\Jonezero}	{J=1\rightarrow 0}

\newcommand{\Sigeoff} {\bar{\Sigma}_{e,{\rm off}}}
\newcommand{\Sigeon} {\bar{\Sigma}_{e,{\rm on}}}
\newcommand{\Sigf} {\bar{\Sigma}_{f}}
\newcommand{\Sige} {\bar{\Sigma}_{e}}

\newlength{\figwidth}
\countdef\decade=200
\advance\decade by \year
\countdef\hours=201
\advance\hours by \time
\divide\hours by 60
\countdef\mins=202
\advance\mins by \hours
\multiply\mins by 60
\multiply\hours by 100
\countdef\miltime=203
\advance\miltime by \hours
\advance\miltime by \time
\advance\miltime by -\mins

\begin{document}

\title{A Virialized Filamentary Infrared Dark Cloud}

\author{Audra K. Hernandez}
\affil{Department of Astronomy, University of Wisconsin, Madison, WI, USA;\\ hernande@astro.wisc.edu}
\author{Jonathan C. Tan}
\affil{Departments of Astronomy \& Physics, University of Florida, Gainesville, FL 32611, USA}
\author{Jouni Kainulainen}
\affil{Max-Planck-Institute for Astronomy, K\"onigstuhl 17, 69117 Heidelberg, Germany}
\author{Paola Caselli}
\affil{School of Physics \& Astronomy, University of Leeds, Leeds, LS2 9JT, UK}
\author{Michael J. Butler}
\affil{Department of Astronomy, University of Florida, Gainesville, FL 32611, USA}
\author{Izaskun Jim\'enez-Serra}
\affil{Harvard-Smithsonian Center for Astrophysics, 60 Garden St., 02138, Cambridge, MA, USA}
\author{Francesco Fontani}
\affil{INAF - Osservatorio Astrofisico di Arcetri, Largo Enrico Fermi 5, I - 50125 Firenze, Italia}

\begin{abstract}
The initial conditions of massive star and star cluster formation are
expected to be cold, dense and high column density regions of the
interstellar medium, which can reveal themselves via near, mid and
even far-infrared absorption as Infrared Dark Clouds
(IRDCs). Elucidating the dynamical state of IRDCs thus constrains
theoretical models of these complex processes. In particular, it is
important to assess whether IRDCs have reached virial equilibrium,
where the internal pressure balances that due to the self-gravitating
weight of the cloud plus the pressure of the external
environmental. We study this question for the filamentary IRDC
G035.39-00.33 by deriving mass from combined NIR \& MIR extinction
maps and velocity dispersion from $\ceto$ (1-0) \& (2-1) line
emission. In contrast to our previous moderately super-virial results
based on $\cothree$ emission and MIR-only extinction mapping, with
improved mass measurements we now find that the filament is consistent
with being in virial equilibrium, at least in its central parsec-wide
region where $\sim 1000\:M_\odot$ snakes along
several parsecs. This equilbrium state does not require
large-scale net support or confinement by magnetic fields.
\end{abstract}

\keywords{ISM: clouds, dust, extinction --- stars: formation}

\section{Introduction} 
Identified by obscuration of the mid-infrared (MIR) (i.e. $\sim
10\:{\rm \mu m}$) Galactic background, Infrared Dark Clouds (IRDCs)
are likely to be representative of the initial conditions of massive
star and star cluster formation, since their high mass surface
densities ($\Sigma \gtrsim 0.1\gcc$) and densities ($n_{\rm H} \gtrsim
10^{4} {\rm cm}^{3}$) are similar to regions with such star formation
activity (e.g. Teyssier et al. 2002; Rathborne et al. 2006; Tan 2007;
Butler \& Tan 2009, hereafter BT09; Zhang et al. 2009; Ragan et
al. 2009; Butler \& Tan 2012, hereafter BT12).

The kinematics of IRDCs can be measured via their molecular line
emission to determine if they are gravitationally bound and/or in
virial equilibrium. Hernandez \& Tan (2011, hereafter HT11) used
$\thco$(1-0) emission from the Galactic Ring Survey (Jackson et
al. 2006) to measure velocity dispersions in two filamentary IRDCs (F
\& H in the BT09 sample). $\Sigma$ was estimated by averaging two
methods: (1) MIR extinction (MIREX) mapping (BT09), given an assumed
MIR opacity per gas mass and a foreground correction based on an
analytic model of Galactic hot dust emission; (2) $\thco$ emission, given an
assumed abundance of this isotopologue. Cloud mass was then calculated
assuming near kinematic distances to the IRDCs. A filamentary virial
analysis following Fiege \& Pudritz (2000, hereafter FP00) was
performed in several orthogonal strips along the filaments. HT11
concluded surface pressure terms are dynamically important, suggesting
that the filaments have not yet reached virial equilibrium.

Here we revisit this analysis for one of the filaments, G035.30-00.33
(H in the BT09/BT12 sample), which is 2.9~kpc distant with a few
thousand solar masses of material spread over about 4~pc (projected) at its
northern end. Although there are several 24~$\rm \mu m$ sources seen
towards the filament (Carey et al. 2009), which are likely to be
embedded protostars, most of the region appears MIR dark and
starless. For our kinematic measurements, we use IRAM 30m observations
of $\ceto$ (1-0) and (2-1). The results of these and other molecular
line observations are being presented in a series of papers on the
formation and evolution of this filamentary IRDC. Paper I
(Jim\'enez-Serra et al. 2010) presented maps of SiO, CO, $\thco$ and
$\ceto$. Widespread SiO emission was observed, perhaps suggesting the
presence of large-scale shocks that may have been involved in forming
the filament. Paper II (Hernandez et al. 2011) compared the $\ceto$
and BT12 MIREX maps, showing CO suffers widespread gas phase depletion
in the IRDC by factors of up to $f_D^\prime \simeq 5$, i.e. just 1 in
5 CO molecules remain in the gas phase. Paper III (this {\it Letter})
performs a filamentary virial analysis of the IRDC. Paper IV (Henshaw
et al. 2012) studies the detailed kinematics of the filament and its
surroundings, including analysis of the dense gas tracer $\rm N_2H^+$.


\section{Mass Surface Density from Extinction Mapping}\label{S:ext}

\begin{figure*}[!tb]
\begin{center}
\includegraphics[height=4.7in, angle=0, trim=6cm 2cm 2cm 0]{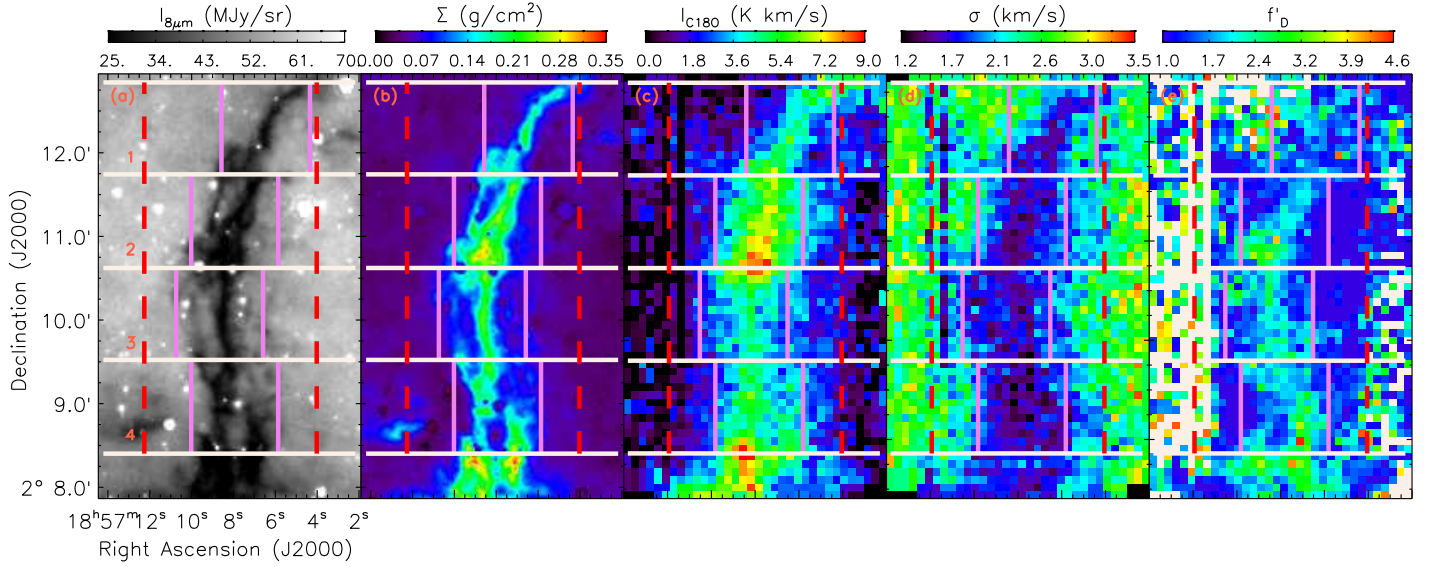} 
\end{center}
\caption{
\small 
IRDC H (G035.30-00.33). {\it Left to Right --- (a):} {\it Spitzer}
GLIMPSE IRAC 8~$\mu$m image. Following HT11, four E-W strips (1-4 from
N to S) have been drawn across the filament. Red dashed vertical lines
delineate either side of the ``Outer Filament''. Magenta solid vertical lines
define the ``Inner Filament'' regions, each half the width of the
Outer Filament and centered on the center of mass of the strip. The
image has 1.2\arcsec\ pixels and PSF with 2\arcsec\ FWHM. {\it (b):}
Total mass surface density, $\Sigma$, derived from (a) using the MIREX
mapping method (BT12) with low mass surface density normalization set
by NIR extinction mapping (KT12). {\it (c):} $\ceto$(2-1) emission
integrated over velocities from $40-50 \kms$ (HT11), a pixel scale of
5\arcsec, and a beam size of 11\arcsec. {\it (d):} Total 1D velocity
dispersion, $\sigma$, derived from the $\ceto$(2-1) $40-50 \kms$
spectra. {\it (e):} Normalized CO depletion factor, $f_D^\prime$, for
regions with $\Sigma_{\rm C18O}\geq 0.005\: \gcc$ (see text).  }
\label{fig:5paneldata}
\end{figure*}

We use the MIREX $\Sigma$ maps of BT12 combined with near infrared
(NIR) (i.e. J,H,K [UKIRT Infrared Deep Sky Survey, Lawrence et
  al. 2007]) extinction maps (Kainulainen et al. 2011a) to yield a
final map that has been presented by Kainulainen \& Tan (2012,
hereafter KT12). The MIREX map is derived from the 2$\arcsec$
resolution {\it Spitzer} IRAC 8~$\rm \mu m$ (Galactic Legacy Mid-Plane
Survey Extraordinaire [GLIMPSE]; Benjamin et al. 2003) image
(Figs. \ref{fig:5paneldata}a \& b).
A MIR (band and background spectrum weighted) dust opacity per unit
total mass of $\kappa_{\rm 8 \mu m}=7.5\:{\rm cm^{2}\:g^{-1}}$ was
adopted, consistent with the moderately-coagulated thin ice mantle
dust model of Ossenkopf \& Henning (1994, hereafter OH94) and a
gas-to-(refractory-component-)dust mass ratio of 156 (Draine \& Lee
1984). We estimate that the uncertainties associated with these
choices are $\sim$30\%, including those due to the grain composition
and size distribution (e.g. the OH94 uncoagulated thin ice mantle
model has opacities 0.83 times smaller; the maximally coagulated model
1.11 times larger; the bare grain, uncoagulated model 0.67 times
smaller) and the gas-to-dust mass ratio (e.g. Draine (2011) estimates
a ratio of 141 from depletion studies, 0.90 times smaller than our
adopted value). Bright MIR sources lead to ``holes'' in the MIREX
map. These cover just a small fraction of the projected area of the
IRDC. They are excluded from the calculation of mean $\Sigma$.

The effect of BT12's use of an empirical foreground estimate (from
observed saturation in independent cores) compared to BT09's
analytic estimate is to increase $\Sigma$, especially in the densest
regions, where it rises by up to a factor of $\sim$3. The average
increase of $\Sigma$ inside the elliptical area of the IRDC defined by
Simon et al. (2006) is a factor of 1.8.

In regions of low $\Sigma$ near the edge of the cloud, the MIREX maps
underestimate $\Sigma$ since the technique is based on relative
extinction of dense regions compared to the surroundings. This is now
corrected for by combining the MIREX maps with NIR extinction maps
(KT12). In regions where the NIR extinction map has $A_V<10$~mag
($\sim 0.04\: \gcc$) it is used to set the zero point of the MIREX
map. In higher column density regions, only the MIREX map is used, now
with the zero point interpolated from the surroundings. 

\section{$\ceto$ Observations to estimate Velocity Dispersion and CO Depletion}\label{S:CO}

The $\ceto$ $J=1-0$ and $2-1$ transitions were mapped (On-The-Fly)
over a $2'\times4'$ area with central position
$\alpha(J2000)=18^h57^m08^s$, $\delta(J2000)=02^{\circ}10'30"$
($l=35.517^{\circ}$, $b=-0.274^{\circ}$) using the IRAM (Insituto de
Radioastronm\'ia Millim\'etrica) 30m telescope in Pico Veleta, Spain
in August \& December 2008 with a combination of the ABCD (for
$\ceto$(1-0)) and HERA ($\ceto$(2-1)) receivers (see Paper II for more details).
At $\sim 110$ GHz ($\Jonezero$), the beam size is 22$\arcsec$, while
at $\sim 220$ GHz ($\Jtwoone$) it is 11$\arcsec$. Data were calibrated
using GILDAS (CLIC).
Spectra were regridded to a velocity resolution of $0.2 \kms$. 

The $\ceto$ emission from the IRDC is between velocities $40-50 \kms$
(HT11). The $\ceto$(2-1) integrated intensity map is shown in Figure
\ref{fig:5paneldata}c. Standard analysis methods (correcting for
excitation temperature and optical depth variations across the
position-velocity cube) were used to derive the distribution of
CO-emitting gas as a function of velocity (Paper II) and thus a map of
total (thermal $+$ nonthermal) 1D velocity dispersion, $\sigma$,
assuming mean particle mass $\mu = 2.33 m_p$ (i.e. $n_{\rm He}=0.2
n_{\rm H_2}$) and temperature of 15~K (Figure
\ref{fig:5paneldata}d). We assume $\sigma$ is measured to about 10\%
accuracy. Note that the IRDC is clearly revealed as a region of
relatively low $\sigma$. Mass surface density assuming a constant
$\ceto$ abundance was derived via $\Sigma_{\rm C18O}=7.65\times
10^{-2} (N_{\rm C18O}/10^{16}{\rm cm^{-2}})\:{\rm g\:cm^{-2}}$ (Paper
II).

The ratio of mass surface density from NIR and MIR extinction,
$\Sigma$ (smoothed to the pixel scale of the CO data), and that from
assuming a constant $\ceto$ abundance, $\Sigma_{\rm C18O}$ defines the
CO depletion factor $f_D \equiv \Sigma/\Sigma_{\rm C18O}$ (averaged
along a sight line through the cloud). To examine relative changes in
depletion, following Paper II, we define a normalized depletion factor
$f_D^\prime = B f_D$, where $B$ is a scaling factor so that
$f_D^\prime$ is on average equal to 1 in regions with
$0.01<\Sigma/\gcc <0.03$. We find $\rm B=0.71$ compared to the Paper
II value of $1.47$ (Case 1 [i.e. no envelope subtraction] HiRes) due
to changes in the $\Sigma$ map, discussed above. For example, if CO is
undepleted in this region and $\Sigma$ correctly estimated, this would
imply we underestimated the true $\ceto$ abundance by $\sim$30\%. The
map of $f_D^\prime$ is shown in Figure~1e, showing values for wherever
$\Sigma_{\rm C18O}\geq0.005\:\gcc$. Given the improved $\Sigma$ map
due to combination with NIR extinction data, this $f_D^\prime$ map
supercedes that of Paper II: in particular it extends to regions of
lower mass surface density. However, the basic structure is unchanged,
with maximum values of $f_D^\prime\simeq 5$ in the highest column
density cores.

\section{Filamentary Virial Analysis}\label{S:virial}

\subsection{Filament and Envelope Geometry}

We assume the filament is locally cylindrical, with radius $R_f$, and
symmetrically embedded in a cylindrical envelope of outer radius $R_e$
(and inner radius $R_f$). To derive the mean mass surface density of
the filament $\bar{\Sigma}_f$, we need to subtract the contribution
from the overlapping portion of the envelope, $\Sigeon$. For a uniform
density envelope and for negligible net contribution of material
outside $R_e$, we first measure $\bar{\Sigma}_{e,{\rm off}}$, i.e. of
the observed envelope region that is offset in projection from the
filament, by averaging the two parts of each strip that are at
projected distances between $R_f$ and $R_e$ from the filament's
central axis. From simple geometry, the mean mass surface density of
the whole envelope, $\Sige$, is related to $\Sigeoff$ via
\begin{equation}
\bar{\Sigma}_e = \frac{\pi {\rm sin}^2\theta (1-{\rm cos}\theta)}{2\theta - {\rm sin}2\theta} \bar{\Sigma}_{e,{\rm off}},
\label{Sigma_env}
\end{equation}
where ${\rm cos}\theta = R_f/R_e$. Thus, since $\bar{\Sigma}_e = (R_f/R_e) \bar{\Sigma}_{e,{\rm on}} + ([R_e-R_f]/R_e) \bar{\Sigma}_{e,{\rm off}}$, we have
\begin{equation}
\bar{\Sigma}_{e,{\rm on}} = \frac{R_e}{R_f}\bar{\Sigma}_e - \frac{R_e-R_f}{R_f}\bar{\Sigma}_{e,{\rm off}}.
\label{Sigma_env,on}
\end{equation}

As shown in Figure~1, we consider two cases: (1) An ``Outer Filament''
(red lines in Figure \ref{fig:5paneldata}) of diameter 2.05\arcmin\ in
R.A. and for which $R_e=1.5R_f$ (set by the extent of the $\rm
C^{18}O$ observations), i.e. $\theta=0.841$ radians,
$\bar{\Sigma}_e=0.845 \bar{\Sigma}_{e,{\rm off}}$ and
$\bar{\Sigma}_{e,{\rm on}} = 0.768 \bar{\Sigma}_{e,{\rm off}}$. Each
of the four strips is centered at $\alpha(J2000)=18^h57^m08.02^s$ and
is 1.12\arcmin\ wide in Dec., with the strip
2-3 boundary at $\delta(J2000)=02^{\circ}10'35.7\arcsec$. This was the
same size and location of the filament region adopted by HT11.
(2) An ``Inner Filament'' (magenta lines in Figure
\ref{fig:5paneldata}) that we set to have half the Outer Filament
width and to have $R_e=2.0R_f$, i.e. $\theta=1.047$,
$\bar{\Sigma}_e=0.959 \bar{\Sigma}_{e,{\rm off}}$ and
$\bar{\Sigma}_{e,{\rm on}} = 0.918 \bar{\Sigma}_{e,{\rm off}}$. Now
within each strip, the location of the Inner Filament is centered on
the center of mass of the whole strip, so this case follows the
structure of the filament more accurately. The correction factors for
the above two cases were applied for estimating $\Sige$ and $\Sigf$.
This division of the cloud into filament and envelope regions is quite
idealized, but is necessary for comparison with analytic models.

The velocity dispersion of the envelope, $\sigma_e$, is measured from
the average spectra of the E and W envelope ``off'' regions in each
strip. Then the above factors for estimating contributions from an
overlapping envelope are also applied to the $\ceto$ spectra: the
spectrum of the filament is derived from the observed ``on'' spectrum
minus the envelope ``off'' spectrum scaled by 0.768 and 0.918 for
Outer and Inner cases, respectively. This subtracted spectrum is used
to derive the total 1D velocity dispersions in the filament,
$\sigma_f$. For both $\sigma_e$ and $\sigma_f$, only those parts of
the spectra with signal $\geq 1$ standard deviation of the noise
level are used.




For each strip and for both Outer and Inner cases, the properties of
the filament and envelope are listed in Table~\ref{tab:1}. In addition
to the 30\% error we assume for the estimates of the directly observed
values of $\Sigma$ (see \S\ref{S:ext}), we also include an additional
30\% error (added in quadrature) for parameters that depend on the
assumed cylindrical filament geometry, i.e. $\Sige$ and $\Sigf$.
Without direct observational constraints, we further assume the
filament axis is inclined by $i=60^\circ \pm 15^\circ$ to our line of
sight ($90^\circ$ would be in the plane of the sky). This uncertainty
contributes up to $22\%$ uncertainty to quantities such as the mass
per unit length of the filament. The distance to the filament is
assumed to be $2.9\pm0.5$~kpc (Simon et al. 2006).
This uncertainty is equivalent to non-circular motions of $\sim 8
\kms$. 


\begin{deluxetable}{l|ccccc|ccccc|c}
\tabletypesize{\footnotesize}
\tablecolumns{12}
\tablewidth{0pt}
\tablecaption{IRDC Filament Properties}
\tablehead{
\multicolumn{1}{c}{Cloud property} & \multicolumn{5}{c}{Outer Filament (strips $1-4$, Total)} & \multicolumn{5}{c}{Inner Filament (strips $1-4$, Total)} & \multicolumn{1}{c}{$\%$ Error}\\
\multicolumn{1}{c}{} & \multicolumn{5}{c}{($R_f=0.865$~pc, $R_e=1.5R_f$)} & \multicolumn{5}{c}{($R_f=0.432$~pc, $R_e=2.0R_f$)} & \multicolumn{1}{c}{}\\
\colhead{}  &\colhead{$1_o$} & \colhead{$2_o$} & \colhead{$3_o$} & \colhead{$4_o$} & \colhead{$T_o$} & \colhead{$1_i$} & \colhead{$2_i$} & \colhead{$3_i$} & \colhead{$4_i$} & \colhead{$T_i$} & \colhead{}}
\startdata
$\Sigeoff$(E) ($\rm 10^{-2}\:g\:cm^{-2}$) & 3.09 & 2.79 & 3.13 & 4.54 & 3.39 & 4.44 & 4.13 & 3.82 & 3.55 & 3.98 & 30\%\\   
$\Sigeoff$(W) ($\rm 10^{-2}\:g\:cm^{-2}$) & 4.23 & 2.55 & 2.34 & 2.62 & 2.93 & 4.25 & 3.74 & 4.75 & 4.60 & 4.34 & 30\%\\   
$\Sigf+\Sigeon$ ($\rm 10^{-2}\:g\:cm^{-2}$) & 5.58 & 8.12 & 7.18 & 7.77 & 7.17 & 7.31 & 12.5 & 10.3 & 11.6 & 10.4 & 30\%\\ 
$\Sige$ ($\rm 10^{-2}\:g\:cm^{-2}$) & 3.09 & 2.26 & 2.33 & 3.03 & 2.67 & 3.67 & 3.32 & 3.62 & 3.43 & 3.52& 42\%\\ 
$\Sigf$ ($\rm 10^{-2}\:g\:cm^{-2}$) & 2.77 & 6.01 & 5.07 & 5.02 & 4.74 & 3.97 & 9.48 & 6.98 & 8.51 & 7.24& 42\%\\ 
$M_e$ ($M_\odot$) & 361 & 264 & 272 & 354 & 1250 & 287 & 259 & 283 & 268 & 1100 & 55\%\\ 
$M_f$ ($M_\odot$) & 216 & 473 & 395 & 392 & 1480 & 155 & 370 & 272 & 332 & 1130 & 55\%\\ 
$n_{{\rm H},e}$ ($10^3{\rm cm}^{-3}$) & 3.27 & 2.39 & 2.46 & 3.21 & 2.83 & 4.32 & 3.91 & 4.27 & 4.04 & 4.14 & 51\%\\ 
$n_{{\rm H},f}$ ($10^3{\rm cm}^{-3}$) & 2.44 & 5.35 & 4.47 & 4.43 & 4.18 & 7.01 & 16.7 & 12.3 & 15.0 & 12.8 & 51\%\\ 
$m_f$ ($M_\odot\:{\rm pc}^{-1}$) & 199 & 435 & 363 & 360 & 340 & 142 & 340 & 250 & 305 & 259 & 51\%\\ \hline
$\bar{v}_f$ (${\rm km\:s^{-1}}$) & 45.54 & 45.41 & 45.34 & 45.05 & 45.32 & 46.14 & 45.62 & 45.80 & 45.19 & 45.72 & ... \\   
$\sigma_f$ (${\rm km\:s^{-1}}$) & 1.59 & 1.47 & 1.37 & 1.25 & 1.36 & 1.17 & 1.31 & 1.01 & 1.01 & 1.08 & 10\%\\   
$m_{{\rm vir},f}$ ($M_\odot {\rm pc^{-1}}$) & 1180	& 1000 & 872 & 722 & 859 & 636 & 804 & 474 & 478 & 543 & 20\%\\ 
$m_f/m_{{\rm vir},f}$ & 0.169 & 0.434 & 0.416 & 0.498 & 0.395 & 0.224 & 0.423 & 0.528 & 0.638 & 0.478 & 55\%\\ \hline
$P_f$ ($10^{-12}{\rm cgs}$) & 145 & 270 & 196 & 161 & 181 & 224 & 677 & 294 & 361 & 349 & 55\%\\ \hline
$\bar{v}_e$ (${\rm km\:s^{-1}}$) & 45.51 & 44.79 & 44.97 & 44.70 & 45.06 & 45.58 & 44.99 & 44.83 & 44.77 & 45.03 & ...\\   
$\sigma_e$ (${\rm km\:s^{-1}}$) & 1.63 & 1.60 & 1.52 & 1.68 & 1.53 & 1.50 & 1.40 & 1.25 & 1.28 & 1.34 & 10\%\\   
$P_e$ ($10^{-12}{\rm cgs}$) & 204 & 143 & 134 & 211 & 154 & 229 & 180 & 156 & 155 & 174 & 55\%\\
$P_e/P_f$  & 1.41 & 0.528 & 0.681 & 1.31 & 0.853 & 1.02 & 0.265 & 0.532 & 0.430 & 0.499 & 41\%\\ \hline
$t_{s,f}=2R_f/\sigma_f$ (Myr) & 1.06 & 1.15 & 1.23 & 1.35 & 1.24 & 0.723 & 0.646 & 0.837 & 0.837 & 0.783 & 20\%\\
$t_{D,f}=0.16 n_{{\rm H},f,4}^{-1}$ (Myr) & 0.66 & 0.30 & 0.36 & 0.36 & 0.38 & 0.23 & 0.096 & 0.13 & 0.11 & 0.13 & $>51$\%\\
\hline
$f_D$&3.35&2.23&2.68&3.24&2.88&2.68&2.36&2.57&2.65&2.57& $50\%$\\
$f_D^\prime$&2.37&1.57&1.89&2.29&2.04&1.89&1.67&1.81&1.88&1.81&$50\%$\\
\enddata
\label{tab:1}
\end{deluxetable}

\subsection{Comparison to Fiege-Pudritz Models of Virial Equilibrium}

We perform a filamentary virial analysis following FP00, who showed
pressure-confined, non-rotating, self-gravitating, filamentary
(i.e. lengths $\gg$ widths) magnetized clouds
that are in virial equilibrium satisfy
\begin{equation}
\frac{P_e}{P_f} = 1-\frac{m_f}{m_{{\rm vir},f}} \left(1-\frac{{\cal M}_f}{|W_f|}\right).
\label{cylvireqn}
\end{equation}
Here: $P_e = \rho_e \sigma_e^2$ (with $\rho_e=\mu_{\rm H} n_{{\rm
    H},e}$, with mass per H nucleus of $\mu_{\rm H} = 1.4 m_p =
2.34\times 10^{-24}$~g) is the external pressure at the surface of the
filament assumed to be equal to the mean envelope pressure; $P_f =
\rho_f \sigma_f^2$ is the mean total pressure in the filament; $m_f$
is the mass per unit length of the filament; $m_{{\rm vir},f}\equiv 2
\sigma_f^2 / G$ is its virial mass per unit length;
${\cal M}_f$ is its magnetic energy per unit length
(associated with large-scale static magnetic fields providing
either support or confinement); and $W_f=-m_f^2G$ is its gravitational
energy per unit length.

\begin{figure*}[!tb]
\begin{center}$
\begin{array}{c}
\includegraphics[width=4.5in,angle=0]{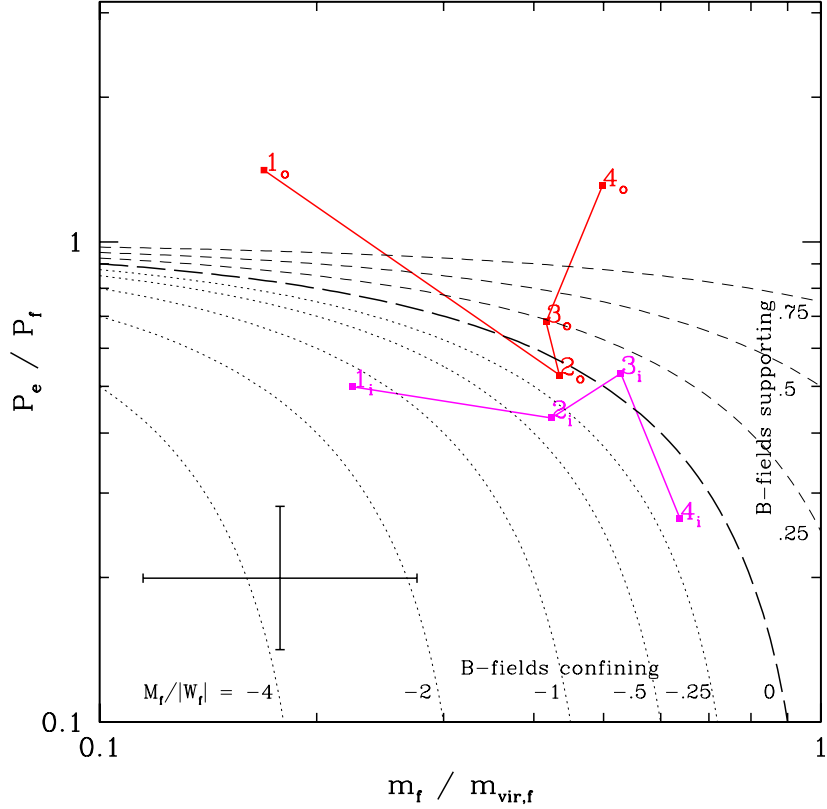} 
\end{array}$
\end{center}
\caption{
Ratio of surface to internal pressure, $P_e/P_f$, versus ratio of mass
per unit length to virial mass per unit length, $m_f/m_{{\rm vir},f}$,
for strips 1 to 4 of IRDC H. Red line and points labelled $\#_o$ show
the ``Outer Filament''. Magenta line and points labelled $\#_i$ show
the ``Inner Filament''. The error bars in the lower-left show typical
uncertainties. The smooth curves show the conditions satisfied by
equation~(\ref{cylvireqn}) for confining magnetic fields with ${\cal
  M}_f/|W_f|<0$ (dotted lines), supporting magnetic fields with ${\cal
  M}_f/|W_f|>0$ (dashed lines), and no magnetic fields, i.e. ${\cal
  M}_f/|W_f|=0$ (long-dashed line). The results for the Inner
Filament
indicate a dynamical state consistent with
magnetically-neutral virial equilibrium.
}
\label{fig:fiege}
\end{figure*}

We can measure $P_e/P_f$ and $m_f/m_{{\rm vir},f}$ for the
Outer and Inner Filament cases for each strip (Table~\ref{tab:1}). No
information is currently available for ${\cal M}_f$. In
Figure~\ref{fig:fiege} we thus compare the observed filament
properties to a series of models that vary the contribution of
magnetic support (${\cal M}_f/|W_f|>0$ from poloidally-dominated
B-fields) or confinement (${\cal M}_f/|W_f|<0$ from
toroidally-dominated B-fields).

Figure~\ref{fig:fiege} indicates that as one zooms from the Outer to
Inner Filaments, $P_e/P_f$ decreases and $m_f/m_{{\rm vir},f}$ tends
to increase. Within the uncertainties, the Inner Filament (and much of
the Outer) is consistent with the model of virial equilibrium not
requiring any magnetic support or confinement (${\cal
  M}_f/|W_f|=0$). Stated another way, for the Inner Filament the
internal pressures are factors of a few greater than external. Such
pressure enhancements are consistent with those expected from
self-gravity given the observed masses and velocity dispersions. Note
that the relative positions of the strips in Figure~\ref{fig:fiege}
are more accurately known than the error in any one position, since
systematic uncertainties are shared. These relative positions,
especially that of strip 4 with respect to 2 and 3, further
support the hypothesis that much of the filament is in virial
equilibrium with a common, small degree of support from large-scale
B-fields.

We have tested the sensitivity of our results to our adopted method of
envelope subtraction: if no correction to filament properties due to
the presence of an overlapping envelope is allowed for, then for the
Inner Filament we derive $\sigma_f$ to be 12\% lower, $P_e/P_f$ to be
23\% lower, and $m_f/m_{{\rm vir},f}$ to be 30\% higher. These values
are still consistent with virial equilibrium without needing
large-scale magnetic support or confinement.


Our results are different from those of HT11, who concluded that, for
the Outer Filament, the surface pressures were all greater than the
internal ones. For Outer Filament strips 1 to 4, they found $P_e/P_f =
3.06, 1.74, 3.32, 5.25$. Our values are factors of 0.46, 0.30, 0.21,
0.25 smaller, respectively. For these strips HT11 found $m_f/m_{{\rm
    vir},f} = 0.0789, 0.161, 0.269, 0.367$. Our values are factors of
2.1, 2.7, 1.5, 1.4 larger, respectively. These differences are mostly
due to our use of improved $\Sigma$ estimation methods. The BT12 MIREX
map improves the estimate of MIR foreground intensity, which increases
the derived values of $\Sigma$ in the denser parts of the
filament. Furthermore, we now no longer use CO line intensities to
derive $\Sigma$: HT11 used $\thco$ line intensities as one method and
then averaged with the BT09 MIREX value. The $\thco$-derived values
were generally lower than the BT09 MIREX values, which we now
attribute as due to CO depletion (Paper II). Thus our estimates of
$\Sigma$ and mass have now increased by factors of a few
compared to HT11, causing $P_e/P_f$ to decrease and
$m_f/m_{{\rm vir},f}$ to increase. 
Our present study uses $\ceto$ to measure velocity dispersions, which
is less prone to being affected by optical depth effects than the
$\thco$ used by HT11 (although HT11 did correct for optical
depth). For the Outer Filament strips 1 to 4, HT11 found $\sigma_f =
1.52, 1.34, 1.03, 0.995\:{\rm km\:s^{-1}}$. Our $\ceto$-derived values
are factors 1.05, 1.10, 1.33, 1.26 times larger. The higher angular
resolution of the $\ceto$(2-1) observations, allow us to probe to
smaller scales, i.e. the narrower Inner Filament case, where line
widths are seen to decrease by factors of about 0.8 compared to the
Outer Filament (Figure 1d). Therefore the overall velocity dispersions
adopted here for the filament are quite similar to those of HT11 and
the differences in our results are thus rather caused by the higher
$\Sigma$'s now derived.

\section{Discussion}\label{S:discussion}

There is some evidence that the IRDC H filament may have formed
recently from converging flows of molecular gas. J\'imenez-Serra et
al. (2010, Paper I) reported widespread, more than parsec-scale SiO
emission from the filament, which may have resulted from large-scale
shocks with speeds of at least several km/s. Henshaw et al. (2012,
Paper IV) studied the kinematics of the region and found evidence for
the formation of the main filament via the merging flow of surrounding
filaments observed in $\ceto$ (with typical densities of $n_{\rm
  H,flow}\sim 10^3\:{\rm cm^{-3}}$). They find that the line of sight relative
velocity between each component of the merging flow and the main
filament is $v_{\rm flow}\sim 3\:{\rm km\:s^{-1}}$.

If the main filament has formed relatively recently, is this
consistent with its observed state of near virial equilibrium? It
should take at least a signal crossing time, $t_{s,f} = 2R_f/\sigma_f$
for a region of the filament to settle into virial equilibrium. These
times are $\sim 0.8$~Myr for the Inner Filament
(Table~\ref{tab:1}). To form the filament from two converging flows takes
a time $t_{{\rm form},f}= 0.236 (m_{f}/100 M_\odot\:{\rm
  pc}^{-1})/[(R_f/{\rm pc})(v_{\rm flow}/3\:{\rm km\:s^{-1}}) (n_{{\rm
      H,flow}}/10^3\:{\rm cm^{-3}})]$~Myr. Applying this to the
average properties of the Inner Filament ($R_f=0.432$~pc, $m_f =
259\:M_\odot$), yields $t_{{\rm form},f}= 1.4$~Myr (see also Paper
IV). Thus, even in the scenario of a recently formed filament, enough
time should have elapsed for it to have settled into virial equilibrium.

The widespread CO depletion observed in Paper II and here (Figure 1e),
also constrains the age of the filament. The CO depletion time is
$t_{D,f}=0.16 (n_{{\rm H},f}/10^4{\rm cm^{-3}})^{-1}$~Myr, assuming a
sticking probability of order unity (Tielens \& Allamandola
1987). Table~\ref{tab:1} lists $t_{D,f}$ for the different strips. For
the Inner Filament, these are relatively short, $\sim0.1$~Myr, which
thus provides a lower limit for its age. 

The implications of virial equilibrium of self-gravitating IRDC
filaments are profound. If this result for G035.39-00.33 applies more
generally to IRDCs, then it indicates that the {\it initial
  conditions} for star formation, including the cores that form
massive stars and the clumps that form star clusters, are created from
environments where approximate pressure equilibrium has been
established. The value of this pressure is set by the self-gravitating
weight of the larger scale cloud, i.e. the IRDC, which dominates over
the pressure of its surrounding environment (unlike for most clumps in
GMCs: e.g., Bertoldi \& McKee 1992; Kainulainen et al. 2011b). This
would confirm a basic assumption of the initial conditions of the Core
Accretion model of massive star formation of McKee \& Tan (2002, 2003)
and is also expected under the scenario of Equilibrium Star Cluster
Formation (Tan, Krumholz \& McKee 2006). The fact that our results
differ from those of HT11, highlights the importance of improved
estimates of masses and surface pressures of IRDCs. These effects may
help explain other reported discrepancies between dynamical
(i.e. virial) masses and true masses (e.g. Battersby et al. 2010).
Measurement of large scale magnetic field strengths and geometries in
IRDCs would help to further constrain the models.

\acknowledgments We thank Bruce Draine and an anonymous referee for
helpful comments and discussions that improved the paper. JCT
acknowledges support from NSF CAREER grant AST-0645412; NASA
Astrophysics Theory and Fundamental Physics grant ATP09-0094; NASA
Astrophysics Data Analysis Program ADAP10-0110. AKH acknowledges 
support from HST grants HST-GO-112275 and HST-GO-12276 to Bart Wakker 
at the University of Wisconsin.  Finally, we would like to acknowledge the IRAM 
facilities and staff.

\end{document}